\documentclass[conference]{IEEEtran}
\usepackage[dvips]{graphicx}
\usepackage{amsmath,amsthm,amsfonts}
\usepackage{amssymb}
\usepackage{hyperref}
\usepackage{verbatim}
\usepackage{mathrsfs}
\usepackage{comment}
\usepackage{ulem}

\newcommand{\cP}{{\mathscr P}}
\newcommand{\cPbar}{\overline{\cP}}
\newcommand{\cX}{{\mathscr X}}
\newcommand{\cY}{{\mathscr Y}}

\newcommand{\cB}{{\mathscr B}}
\newcommand{\bR}{\mathbb{R}}

\newcommand{\be}{\begin{equation}}
\newcommand{\ee}{\end{equation}}

\newcommand{\e}{({\rm e})}
\newcommand{\m}{({\rm m})}

\renewcommand{\d}{{\rm d}}

\newcommand{\ba}{\begin{align}}

\renewcommand{\em}{\it}

\begin{document}

\title{Information-geometrical  characterization of statistical models which are \\
statistically equivalent to probability simplexes}

\author{\IEEEauthorblockN{\large Hiroshi Nagaoka}
\IEEEauthorblockA{Graduate School of Informatics and Engineering\\
The University of Electro-Communications\\
Chofu, Tokyo 182-8585, Japan\\
Email: nagaoka@is.uec.ac.jp}
}

\maketitle

\begin{abstract}
The probability simplex is the set of all probability distributions on a finite set and is 
the most fundamental object in the finite probability theory. 
In this paper we give a characterization of 
statistical models on finite sets which are statistically equivalent  
to probability simplexes in terms of $\alpha$-families including exponential families and 
mixture families.  The subject has a close relation to some fundamental aspects of information geometry such as $\alpha$-connections and autoparallelity. 
\end{abstract}


%
\IEEEpeerreviewmaketitle 

\section{An introductory example}

Let $\cX = \{0, 1, 2\}$ and 
let $M =\{ p_\lambda \, |\, 0 < \lambda <1\}$  
be the set of probability distributions on $\cX$ of the form
\[  p_\lambda = (p_\lambda (0), p_\lambda (1), p_\lambda (2)) = 
\left(\lambda, (1-\lambda)/2, (1-\lambda)/2\right). 
\]
The statistical model $M$ has the following three properties. Firstly, it is a mixture family since 
\[ p_\lambda = \lambda\, (1, 0, 0) + (1-\lambda)\, (0, 1/2, 1/2). 
\]
Secondly, it is an exponential family since 
\[
\log p_\lambda = \theta F - \psi (\theta), 
\]
where $\theta = \log (2\lambda /(1-\lambda))$, $(F(0), F(1), F(2)) = (1, 0, 0)$ and 
$\psi (\theta ) = - \log (1-\lambda)/2 =\log (2+e^\theta)$.  Lastly, 
$M$ is statistically equivalent to the 1-dimensional open probability simplex 
$\cP_1 = \{(\lambda, 1-\lambda)\,|\, 0 < \lambda < 1\}$ 
in the sense that there exist a channel $V$ from $\{0, 1\}$ to $\cX$ and 
a channel $W$ from $\cX$ to $\{0, 1\}$ such that $M$ is the set of 
output distributions of $V$ for input distributions in $\cP_1$ 
and that $V$ is invertible by $W$.  The matrix representations of these channels are 
given by
\[
V = \left[
\begin{matrix}
1 & 0  \\
0 & 1/2 \\
0 & 1/2
\end{matrix}
\right], 
\quad
W = \left[
\begin{matrix}
1 & 0 & 0 \\
0 & 1 & 1
\end{matrix}
\right]. 
\]
Note that the invertibility $W V = I$ holds. 

Our aim is to show the equivalence between the first two properties and 
the last one. 

\section{Statement of the main result}

We begin with giving some basic definitions which are necessary to state our problem.  

For an arbitrary finite set $\cX$, let $\cPbar (\cX)$ and $\cP(\cX)$ be 
the sets of probability distributions and  of strictly positive probability distributions on $\cX$; 
\begin{align*}
 \cPbar (\cX) &:= \{p \,|\, p : \cX \rightarrow [0, 1], \; 
\sum_x p(x) =1\} \\
 \cP(\cX) &:= \{p \,|\, p : \cX \rightarrow (0, 1), \; 
\sum_x p(x) =1\}. 
\end{align*}
In particular, let for an arbitrary positive integer $d$ 
\begin{align*}
{\cPbar}_d &:= \cPbar (\{0, 1, \ldots , d\}) \\
{\cP}_d &:= \cP(\{0, 1, \ldots , d\}), 
\end{align*}
which we call the $d$-dimensional (closed and open) probability simplexes. 

A mapping $\Gamma : \cPbar (\cX) \rightarrow  \cPbar (\cY)$, 
where $\cX$ and $\cY$ are finite sets, is called a {\em Markov map} when 
there exists a channel $W(y|x)$ from $\cX$ to $\cY$ such that, for any $p\in \cPbar (\cX) $, 
\[
 \Gamma (p) = \sum_x W(\,\cdot\, | x) p(x).
\]
i.e.,  $\Gamma (p)$ is the output distribution of the channel $W$ 
corresponding to 
the input distribution $p$.  Note that a Markov map is 
affine; $\Gamma (\lambda p + (1-\lambda ) q) = \lambda \Gamma (p) + (1-\lambda ) \Gamma (q)$ for $\forall p, q\in \cPbar (\cX)$ and $0\leq \forall \lambda \leq 1$. 

Let $M$ and $N$ be smooth submanifolds (statistical models) of $\cP (\cX)$ and $\cP (\cY)$, respectively. When there exist a pair of Markov maps $\Gamma : \cPbar (\cX) \rightarrow  \cPbar (\cY)$ and $\Delta : \cPbar (\cY) \rightarrow  \cPbar (\cX)$ such that their restrictions $\Gamma |_M$ and $\Delta |_N$
are bijections between $M$ and $N$ and are the inverse mappings of each other, we say that $M$ and $N$ are {\em Markov equivalent} or 
{\em statistically equivalent} and wite as
$M \simeq N$.  

The aim of this paper is to give a characterization of  statistical models which are statistically equivalent to probability simplexes.  The main result is  as follows. 

\medskip

\noindent {\bf Theorem 1}  \,
For an arbitrary smooth submanifold $M$ of $\cP (\cX) $, the following conditions are mutually equivalent. 
\begin{quote}
\begin{itemize}
\item[(i)] $M\simeq \cP_d$, where $d= \dim M$. 
\item[(ii)] $M$ is an exponential family and is a mixture family. 
\item[(iii)] $\exists \alpha \neq \exists \beta$, $M$ is an $\alpha$-family and 
is an $\beta$-family. 
\item[(iv)] $\forall \alpha$, $M$ is an $\alpha$-family. 
\end{itemize}
\end{quote}

\medskip

Explanation of exponential family, mixture family and $\alpha$-family for 
arbitrary 
$\alpha\in \bR$ as well as the proof of the theorem 
will be presented in subsequent sections.  Here we only give a few remarks on 
condition (i).  Firstly, (i) is equivalent to the condition that $\exists d', \; M\simeq \cP_{d'}$, since 
if $M\simeq \cP_{d'}$ then $M$ and  $\cP_{d'}$ must be 
diffeomorphic,  so that $ \dim M = \dim  \cP_{d'} = d'$. Secondly, (i) is equivalent to 
the condition $\overline{M} \simeq \overline{\cP}_{d}$, where $\overline{M}$ denotes 
the topological closure of $M$, and means that $\overline{M}$ 
is the set of output distributions of an invertible (erro-free) channel. 

\section{Some facts about condition {\rm (i)}}
\label{sec_condition(i)}

From the definition of the relation $\simeq$, condition (i) implies that 
there exist $\Gamma: \cPbar (\cX) \rightarrow {\cPbar}_d$ and 
$\Delta : {\cPbar}_d \rightarrow {\cPbar} (\cX)$ satisfying 
$\Gamma \circ \Delta = {\rm id}$ (the identity on ${\cPbar}_d$).  
Let $\{q_0, q_1, \ldots , q_d\} \subset \cPbar (\cX)$ be defined by 
\be
 \Delta (\delta_i ) = q_i, \;\; \forall i \in \{0, 1, \ldots , d\}, 
 \label{Delta}
 \ee
 where $\delta_i$ is the delta distributions on $\{0, 1, \ldots , d\}$  concentrated on $i$. 
Then it is easy to see, as is shown in Lemma~9.5 and its ``Supplement" 
of \cite{Chentsov} where our $\Delta$ is called a 
{\em congruent embedding} (of ${\cPbar}_d$ into $\cPbar (\cX)$), that 
the supports 
$A_i := {\rm supp}\, (q_i) $ constitute a partition of 
$\cX$  in the sense that
\vspace{-2mm}
\be 
\vspace{-2mm}
A_i \cap A_j = \phi\;\;\text{if}\;\; i\neq j, 
\;\;\text{and}\;\; \bigcup_{i=0}^d A_i = \cX, 
 \label{partition}
 \vspace{-2mm}
\ee
\vspace{-2mm}
and the left inverse $\Gamma$ of $\Delta$ is represented as 
\be
\Gamma (p) = \sum_{i=0}^d  p(A_i) \, \delta_i, \;\; \forall p\in \cPbar (\cX), 
 \label{Gamma}
\ee
where $p(A_i) := \sum_{x\in A_i} p(x)$. 
In addition, condition (i) implies $M=\Delta ({\cP}_d):= 
\{ \Delta (\lambda)\,|\, \lambda\in {\cP}_d\}$, so that from \eqref{Delta} we have
\be
M = \left\{
\sum_{i=0}^d \lambda_i q_i\;\Big|\; (\lambda_0, \ldots , \lambda_d)\in {\cP}_d
\right\}.
\label{M:cond(i)}
\ee
Conversely, if a statistical model $M \subset \cP (\cX)$ is represented in 
the form \eqref{M:cond(i)} by a collection of $d+1$ distributions $\{q_i\}$ on $\cX$ whose supports $\{A_i\}$ constitute a partition of $\cX$, then we see that $M$ satisfies condition (i) by 
defining $\Delta$ and $\Gamma$ by \eqref{Delta} and \eqref{Gamma}.  Thus 
a necessary and sufficient condition for (i) is obtained, which will be used in later arguments to prove 
the theorem. 

\section{$\alpha$-family, {\rm e}-family and {\rm m}-family}
\label{section:alpha_e_m_family}

Following the way developed in \cite{NA} (see also 
\cite{Amari:springer, AN}), 
we give the definition of $\alpha$-family, which includes that of {\em exponential family} and 
 {\em mixture family} as special cases. 
 
 For an arbitrary $\alpha\in\bR$, define a function $L^{(\alpha)} : \bR^+ 
 (= (0, \infty)) 
 \rightarrow 
 \bR$ by\footnote{%
$L^{(\alpha)} (u) $ can be replaced with $a L^{(\alpha)} (u) + b$ by arbitrary constants $a\neq 0$ and $b$, possibly depending on $\alpha$.  In \cite{Amari:springer, AN, NA}, these constants are properly chosen so that 
the $\pm \alpha$-duality and the 
limit of $\alpha\rightarrow 1$ can be treated in a convenient way. 
 }
 \be
 L^{(\alpha)} (u) = \left\{
 \begin{array}{cc} 
 u^{\frac{1-\alpha}{2}} & (\alpha \neq 1) \\
 \log u & (\alpha =1). 
 \end{array}
 \right.
 \label{def_Lalpha} 
 \ee
The function $L^{(\alpha)}$ is naturally extended to a mapping $(\bR^+)^{\cX} \rightarrow 
 {\bR}^{\cX}$  ($f\mapsto L^{(\alpha)} (f)$) 
 by 
 \be
  \left(L^{(\alpha)} (f)\right) (x) = L^{(\alpha)} (f(x)). 
  \label{extend_Lalpha}
\ee

For a submanifold $M$ of $\cP (\cX)$, its {\em denormalization} $\tilde{M}$ 
is defined by 
\be
 \tilde{M} := \left\{ \tau p\, |\, p\in M \;\;\text{and}\;\; \tau \in \bR^+\right\},
\label{def_denormalization}
\ee
where $\tau p$ denotes the function $\cX\ni x \mapsto \tau p(x) \in \bR^+$. 
The denormalization is an extended manifold obtained by relaxing the normalization 
constraint $\sum_x p(x) =1$. 
Obviously, $\tilde{M}$ is a submanifold of $\widetilde{\cP} (\cX)$, 
and $\widetilde{\cP} (\cX) = (\bR^+)^{\cX}$ 
is an open subset of 
$ {\bR}^{\cX}$.  
When the image 
\[L^{(\alpha)} (\tilde{M}) = \left\{ L^{(\alpha)} (\tau p)\;\Big|\; 
p\in M \;\;\text{and}\;\; \tau \in \bR^+\right\}\]
forms an open subset of an affine subspace, say $Z$,  of  $\bR^{\cX}$, 
$M$ is called an {\em $\alpha$-family}.  In this paper, 
it is assumed for simplicity that $M$ is 
maximal in the sense that 
\be
L^{(\alpha)} (\tilde{M}) =  Z \cap L^{(\alpha)} \left( (\bR^+)^{\cX}\right).
\label{maximal_alpha1}
\ee
Since it follows from  the definition \eqref{def_Lalpha} of $L^{(\alpha)}$ that 
\[
L^{(\alpha)} \left( (\bR^+)^{\cX}\right) = 
\left\{
\begin{array}{cc} 
  (\bR^+)^{\cX} & (\alpha \neq 1) \\
 {\bR}^{\cX} & (\alpha =1), 
 \end{array}
 \right. 
\]
\eqref{maximal_alpha1} is written as
\be
L^{(\alpha)} (\tilde{M}) 
 = 
\left\{
\begin{array}{cc} 
Z \cap  (\bR^+)^{\cX} & (\alpha \neq 1) \\
Z  & (\alpha =1). 
 \end{array}
 \right. 
\label{maximal_alpha2}
\ee
Note that, as is pointed out in section~2.6 of \cite{AN}, an affine subspace $Z$ satisfying \eqref{maximal_alpha2} must be a linear subspace when $\alpha\neq 1$. 
Note  also that $\cP (\cX)$ is an $\alpha$-family for $\forall \alpha\in \bR$, corresponding to the case when $Z = \bR^{\cX}$. 

When $\alpha =1$, the notion of $\alpha$-family is equivalent to that of 
exponential family, whose general form is $M = \{p_\theta\, |\, 
\theta=(\theta^1, \ldots , \theta^d) \in \bR^d \}$ such that 
\be
 p_\theta (x) =  \exp\left[ C(x) + \sum_{i=1}^d \theta^i F_i(x) - \psi(\theta)\right], 
 \label{e-family}
\ee
where  $C, F_1, \ldots , F_d$ are functions on $\cX$ 
and $\psi$ is a function on $ \bR^d$ defined by 
\be \psi (\theta) = \log\sum_x  \exp\left[C(x) +\sum_{i=1}^d \theta^i F_i(x)\right] .
\label{normalize}
\ee

When $\alpha =-1$, on the other hand, the notion of $\alpha$-family is equivalent to 
that of mixture family, whose general form is 
$M = \{p_\theta\, |\, 
\theta=(\theta^1, \ldots , \theta^d) \in \Theta \}$
such that 
\be
p_\theta (x) = C(x) +  \sum_{i=1}^d \theta^i\, F_i(x), 
\label{m-family}
\ee
where  $F_1, \ldots , F_d$ are functions on $\cX$ satisfying \\
$\sum_x F_i (x) = 0$ and $\Theta := 
\{\theta\in \bR^d\, |\, 
\forall x, \, p_\theta (x)>0 
\}$.  

When $\alpha \neq 1$, the general form of $\alpha$-family $M = \{p_\theta\, |\, 
\theta=(\theta^1, \ldots , \theta^d) \in \Theta \}$ is 
\begin{align}
p_\theta (x) = \Bigl\{
\sum_{j=0}^d \xi^j (\theta) F_j (x)\Bigr\}^{\frac{2}{1-\alpha}}. 
\end{align}
See \S 2.6 of \cite{AN} for further details.

\section{Proof of {\rm (i)} $\Rightarrow$ {\rm (iv)}}

Assume (i), which implies that there exists a collection of $d+1$ probability distributions 
$\{ q_i\} \subset \cPbar (\cX)$ whose supports $\{A_i\}$ constitute a partition of $\cX$ 
and that $M$ is represented as \eqref{M:cond(i)}.  Then the denormalization $\tilde{M}$ 
is represented as
\be
\tilde{M} = \left\{
\sum_{i=0}^d \lambda_i q_i\;\Big|\; (\lambda_0, \ldots , \lambda_d)\in 
({\bR}^+)^{d+1}
\right\}.
\label{tildeM:cond(i)}
\ee

Let $\alpha$ be an arbitrary real number such that $\alpha\neq 1$. Since  
$L^{(\alpha)} (0) = 0$ in this case,  it follows 
from the disjointness of the supports of $\{q_i\}$ 
that 
\[ 
L^{(\alpha)} \left(\sum_i \lambda_i q_i\right) = \sum_i \lambda_i^{\frac{1-\alpha}{2}} 
L^{(\alpha)}(q_i)
\]
for any $ (\lambda_0,  \ldots , \lambda_d)\in 
({\bR}^+)^{d+1}$. 
From this we have
\begin{align*}
& L^{(\alpha)} (\tilde{M}) \\
&= 
\left\{
\sum_{i=0}^d  \lambda_i^{\frac{1-\alpha}{2}}  L^{(\alpha)}(q_i)
\;\Big|\; (\lambda_0,  \ldots , \lambda_d)\in 
({\bR}^+)^{d+1}
\right\} \\
& = 
\left\{
\sum_{i=0}^d \xi_i L^{(\alpha)}(q_i)
\;\Big|\; (\xi_0, \ldots , \xi_d)\in 
({\bR}^+)^{d+1}
\right\} \\
& = Z \cap  (\bR^+)^{\cX}, 
\end{align*}
where $Z$ is the $(d+1)$-dimensional linear subspace of ${\bR}^{\cX}$ spanned by
$L^{(\alpha)}(q_i)$, $i\in\{0, 1, \ldots , d\}$. This proves that $M$ is an $\alpha$-family 
for any $\alpha \neq 1$. 

Let $\alpha =1$. For any $x\in \cX$, we have
\begin{align*}
L^{(1)} \left(\sum_i \lambda_i q_i\right) (x) &= 
\log  \left(\sum_i \lambda_i q_i (x) \right) \\
&= \log (\lambda_j q_j (x)) \\
&= \log\lambda_j + \log q_j (x) \\
&= \sum_{i} \left(  \log\lambda_i + \log q_i (x)\right) 1_{A_i}(x),
\end{align*}
where $j$ denotes the element of $\{0, 1, \ldots , d\}$ 
such that $x\in A_j$.  Letting $C\in {\bR}^{\cX}$ 
be defined by $C(x) = \sum_i (\log q_i(x)) 1_{A_i} (x)$, 
we have
\begin{align*}
& L^{(1)} (\tilde{M}) \\
&= 
\left\{
C+ 
\sum_{i=0}^d (\log \lambda_i) 1_{A_i}
\;\Big|\; (\lambda_0, \ldots , \lambda_d)\in 
({\bR}^+)^{d+1}
\right\} \\
&= \left\{
C+ 
\sum_{i=0}^d \xi_i 1_{A_i}
\;\Big|\; (\xi_0, \ldots , \xi_d)\in 
{\bR}^{d+1}
\right\}, 
\end{align*}
which is an affine subspace of ${\bR}^{\cX}$. 
This proves that $M$ is a $1$-family (an exponential family). 

The implication (i) $\Rightarrow$ (iv) has thus been proved. 

\section{Equivalence of {\rm (ii), (iii)} and {\rm (iv)}}

The implications (iv) $\Rightarrow$ (ii) $\Rightarrow$ (iii) 
are obvious.  To see  (iii) $\Rightarrow$ (iv), some results of information geometry are invoked. 

\medskip
\noindent
{\em Remark 1:}\quad 
The notion of affine connections appears only in this section. 
Since the implication (ii) $\Rightarrow$ (i) will be proved in the next section without using affine connections (at least explicitly), we do not need them in proving 
the equivalence of the conditions of Theorem~1 except for (iii). 
\medskip

We first introduce some concepts from general differential geometry. 
Let $S$ be a smooth manifold and denote by ${\cal T} (S)$  
the set of smooth vector fields on $S$. Here, by a vector field on $S$ we mean 
a mapping, say $X$, such that $X : S\ni p \mapsto X_p\in T_p (S)$, where 
$T_p (S)$ denotes the tangent space of $S$ at $p$. 
An affine connection on $S$ is represented by a mapping $\nabla : 
{\cal T} (S) \times {\cal T} (S) \ni (X, Y) \mapsto \nabla_X Y \in {\cal T} (S)$, 
which is called a covariant derivative, 
satisfying certain conditions. 
Let $M$ be a smooth submanifold of $S$.  Then 
$\nabla$ is naturally defined on 
${\cal T}(M) \times {\cal T}(M)$, 
so that $\nabla_XY$ is defined for any vector fields on $M$. However, 
the value $\nabla_XY$ in this case is a mapping $M\ni p\mapsto (\nabla_XY)_p \in T_p (S)$ in general 
and is not a vector field on $M$ (i.e., $\nabla_X Y\not\in {\cal T}(M)$) 
unless 
\be
(\nabla_XY)_p 
\in T_p (M),  \;\; \forall p\in M.
\label{autoparallel}
\ee
When \eqref{autoparallel} holds  for 
$\forall X, Y\in {\cal T}(M)$, $M$ is said to be {\em autoparallel} w.r.t.\ 
$\nabla$ or {\em $\nabla$-autoparallel} in $S$.  

Let $\nabla, \nabla'$ and $\nabla''$ be affine connection on $S$ for which there exists a real number $a$ satisfying\footnote{%
For arbitrary affine connections $\nabla$ and $\nabla'$, their affine combination 
$a \nabla + (1-a) \nabla'$ always becomes an affine connection.
}
\begin{equation} \nabla'' = a \nabla + (1-a) \nabla' . 
\label{affine_combination_connections}
\end{equation}
If a submanifold $M$ is $\nabla$-autoparallel and $\nabla'$-autoparallel, then 
it is also $\nabla''$-autoparallel. This implication is obvious from 
$(\nabla''_X Y)_p = a (\nabla_X Y)_p + (1-a) (\nabla'_X Y)_p
$
and the autoparallelity condition \eqref{autoparallel}, which will be invoked later. 

As was independently introduced by \v{C}encov \cite{Chentsov} and Amari 
\cite{Amari:CurvedExp82},
a one-parameter  
family of affince connections, which are called the {\em $\alpha$-connections} ($\alpha \in \bR$), 
are defined on a manifold of probability distributions.
After Amari's notation, 
the $\alpha$-connection is written in the form of affine combination 
\be
 \nabla^{(\alpha)} = \frac{1+\alpha}{2} \nabla^{(1)} + 
\frac{1-\alpha}{2} \nabla^{(-1)}, 
\ee
which implies that
\be
\nabla^{(\gamma)} = 
\frac{\gamma -\beta}{\alpha - \beta} 
\nabla^{(\alpha)} + 
\frac{\alpha - \gamma}{\alpha - \beta} \nabla^{(\beta)}
\label{alpha_beta_gamma}
\ee
for any $\alpha, \beta, \gamma \in \bR$ such that $\alpha\neq \beta$. 

When  a submanifold $M$ of $S$ is autoparallel w.r.t.\ the $\alpha$-connection 
in $S$, we say that $M$ is {\em $\alpha$-autoparallel} in $S$.  
Since \eqref{alpha_beta_gamma}  is of the form \eqref{affine_combination_connections}, 
it follows that if $M$ is $\alpha$-autoparallel and 
$\beta$-autoparallel in $S$ for 
some $\alpha\neq \beta$, then it is $\gamma$-autoparallel in $S$ for all $\gamma\in\bR$. 
On the other hand, 
it was shown in \cite{NA} (see also section~2.6 of  \cite{AN}) that, for any submanifold $M$ of $\cP (\cX)$ and for any real number $\alpha$, $M$ is an $\alpha$-family 
if and only if $M$ is $\alpha$-autoparallel in $\cP (\cX)$. 
  Combination of these two results 
proves  (iii) $\Rightarrow$ (iv). 

\medskip
\noindent
{\em Remark 2:}\quad 
Since the e-connection and the m-connection are dual w.r.t. the Fisher information metric
\cite{Amari:springer, AN, NA}, condition (ii) is a special case of doubly autoparallelity introduced 
by Ohara; see  \cite{Ohara2, Ohara3} and the reference cited there.  It is pointed out in \cite{Ohara3} that the $\alpha$-autoparallelity for all $\alpha$ follows from that for $\alpha = \pm 1$. 

\medskip

\section{Proof of {\rm (ii)} $\Rightarrow$ {\rm (i)}}

Assume (ii), which means that there exist two affine subspaces $Z^{\e}$ and $Z^{\m}$ of 
$\bR^{\cX}$ such that 
\be L^{\e}(\tilde{M}) = \{ \log \mu \,|\, \mu\in \tilde{M}\} = Z^{\e}
\label{def_Ze}
\ee
\be L^{\m}(\tilde{M}) = \tilde{M}  = Z^{\m} \cap ({\bR}^+)^{\cX} ,
\label{def_Zm}
\ee
where $L^{\e}:= L^{(1)}$ and $L^{\m}:= L^{(-1)}$. 
Let $V^{\e}$ and $V^{\m}$ be the linear spaces of translation vectors of 
$Z^{\e}$ and $Z^{\m}$, respectively, so that we have $Z^{\e} = 
f + V^{\e}$ for any $f\in Z^{\e} $ and $Z^{\m} = 
g + V^{\m}$ for any $g\in Z^{\m} $\footnote{%
Actually, $Z^{\m}$ is a linear space as mentioned in section~\ref{section:alpha_e_m_family}, and therefore $Z^{\m} = V^{\m}$. 
}.

\medskip
\begin{description}
\item[{\bf Lemma 1}] \qquad
\  $V^{\e}$ is closed w.r.t.\ 
multiplication of functions; i.e., $a, b\in V^{\e} 
 \Rightarrow  a b\in V^{\e}$, where the product 
 $ab$ is defined by $(ab) (x) = a(x) b(x)$. 
\end{description}

\medskip
\begin{proof}
The map 
\[
\Phi := L^{\e}|_{\tilde{M}} : \tilde{M}\ni \mu \mapsto \log \mu \in 
Z^{\e} 
\]
is a diffeomorphism from $\tilde{M} = Z^{\m} \cap ({\bR}^+)^{\cX}$, which 
is an open subset of $Z^{\m}$, onto $Z^{\e}$.  The differential 
map of $\Phi$ at a point $\mu\in \tilde{M}$ is defined by 
\[ (\d\Phi)_\mu \Bigl(
\frac{d\mu(t)}{dt}\Big|_{t=0}\Bigr) = 
\frac{d}{dt} \Phi ( \mu (t) ) \Big|_{t=0}
\]
for any smooth curve $\mu (t)$ in $\tilde{M}$ and is 
represented as
\[
(\d\Phi)_\mu : V^{\m}\ni f \mapsto \frac{f}{\mu}\in V^{\e}.
\]
This gives a linear isomorphism from $V^{\m}$ onto $V^{\e}$. 
Therefore, for any two points $\mu, \nu\in\tilde{M}$, we can define
\vspace{-1mm}
\[ 
(\d\Phi)_\nu \circ (\d\Phi)_\mu^{-1} : V^{\e}\ni a \mapsto \frac{\mu a}{\nu}\in V^{\e}.
\vspace{-2mm}
\]
This means that, for any $a\in V^{\e}$ and any $\mu, \nu\in\tilde{M}$, we have 
$\frac{\mu a}{\nu}\in V^{\e}$. 
For arbitrary $a\in V^{\e}$ and $\nu\in\tilde{M}$, let us define a map $\Psi_{a, \nu}$  by
\[ 
\vspace{-1mm}
\Psi_{a, \nu} : \tilde{M}\ni \mu \mapsto \frac{\mu a}{\nu}\in V^{\e}. 
\vspace{-2mm}
\]
Then its differential at a point $\mu\in\tilde{M}$ is given by
\[
(\d \Psi_{a, \nu})_\mu : V^{\m}\ni g\mapsto \frac{g a}{\nu} \in V^{\e}.
\]
Composing this map with the inverse of 
\[ (\d \Phi)_\nu 
: V^{\m}\ni g \mapsto \frac{g}{\nu}\in V^{\e}, 
\]
we have
\[
(\d \Psi_{a, \nu})_\mu \circ  (\d \Phi)_\nu^{-1} : 
V^{\e} \ni b \mapsto ab \in V^{\e}.
\]
This proves that $a, b\in V^{\e} \, \Rightarrow ab\in V^{\e}$.
\end{proof}

\begin{description}
\item[{\bf Lemma 2}]  \qquad
$V^{\e}$ contains the constant functions on $\cX$.  
\end{description}

\begin{proof}
From the definition \eqref{def_denormalization} of  $\tilde{M}$, for 
any $\mu\in\tilde{M}$ and any positive constant $\tau = e^c$, we have 
$\tau \mu\in\tilde{M}$.  This implies that  both $\log \mu$ and 
$\log (\tau \mu)$ belong to $Z^{\e}$, and hence the translation 
$\log (\tau \mu) - \log \mu = \log \tau = c$ belongs to $V^{\e}$.  
\end{proof}

These two lemmas state that $V^{\e}$ is a subalgebra of the commutative algebra 
$\bR^{\cX}$ with the unit element $1$ (: the constant function $x\mapsto 1$) 
of $\bR^{\cX}$ contained in $V^{\e}$. From a well known result on such subalgebras\footnote{%
Although 
various mathematical extensions of this result including 
infinite-dimensional and/or noncommutative versions 
are known, the author of the present paper could  find 
no appropriate reference describing the result for 
the finite-dimensional commutative case with an 
elementary proof.  So, we give a proof in the appendix for the readers' sake. 
}
, it is 
concluded that there exists a partition $\{A_i\}_{i=0}^d$ of $\cX$ such that 
\be
V^{\e} = \left\{
\sum_{i=0}^d c_i 1_{A_i}\, \Big|\, (c_0, \ldots , c_d)\in {\bR}^{d+1}\right\}. 
\label{Ve_partition}
\ee
Let an element $p_0$ of $M\, (\subset \tilde{M})$ be arbitrarily fixed. 
Then we have
\be
 Z^{\e} = \log p_0 + V^{\e}. 
\label{Ze_logp_0}
\ee
From \eqref{def_Ze}, \eqref{Ve_partition} and \eqref{Ze_logp_0} and the disjointness of $\{A_i\}$, we have
\begin{align*}
\tilde{M} 
& =  \{\mu\,|\, \log \mu\in Z^{\e} \} \\
& =  \{\mu\,|\, \log \mu -\log p_0 \in V^{\e} \} \\
&=  \Big\{\mu\,\Big|\, 
 \exists (c_0, \ldots , c_d) \in {\bR}^{d+1},  \\
& \hspace{1cm}
\log \mu = \log p_0 + \sum_{i=0}^d c_i 1_{A_i},  
 \Big\}, 
\\
&= \left\{ p_0  \sum_{i=0}^d e^{c_i} 1_{A_i} \, \Big|\, (c_0, \ldots , c_d)\in {\bR}^{d+1}\right\} \\
&=  \left\{ 
\sum_{i=0}^d \lambda_i q_i \, \Big|\, 
(\lambda_0, \ldots , \lambda_d)\in ({\bR}^+)^{d+1}\right\}, 
\end{align*}
where 
\[ q_i := \frac{1}{p_0 (A_i)}\, p_0 1_{A_i}, \quad 
i\in\{0, \ldots , d\}.
\]
Then $\{q_i\}$ are probability distributions on $\cX$ whose 
supports are ${\rm supp}\, (q_i) = A_i$, and
\begin{align*}
M & = \tilde{M} \cap \cP (\cX) \\
&= \left\{ 
\sum_{i=0}^d \lambda_i q_i \, \Big|\, 
(\lambda_0, \ldots , \lambda_d)\in {\cP}_d \right\}. 
\end{align*}
Since this is the same form as \eqref{M:cond(i)}, condition (i) has been derived. 

\section{Conclusion}
We have shown Theorem~1 which gives 
an information-geometrical characterization of statistical models 
on finite sample spaces which are statistically equivalent to open probability 
simplexes $\cP_d$.  
The statistical equivalence (also called the Markov equivalence) to 
probability simplexes played a crucial role in \v{C}encov's pioneering 
work \cite{Chentsov} on information geometry, where the notions 
of Fisher information metric and the $\alpha$-connections were 
characterized in terms of the statistical equivalence. 
The present work shed another light on the relation between the statistical equivalence 
and information geometry. 

\section*{Acknowledgment}
This work was partially supported by JSPS KAKENHI Grant Number 16K00012.

\section*{Appendix}

\noindent 
{\bf Proposition}  \, 
Let $\cX$ be a finite set and $V$ be 
a subalgebra of $\bR^{\cX}$ containing the constant functions. 
Then 
there exists a partition $\{A_i\}_{i=1}^n$ of $\cX$ such that 
\be
V = \left\{
\sum_{i=1}^n c_i 1_{A_i}\, \Big|\, (c_1, \ldots , c_n)\in {\bR}^{n}\right\}. 
\label{V_partition}
\ee

\medskip
\begin{proof}
Let
\be
\cB := \left\{ f^{-1} (\lambda)\, |\, \lambda\in \bR, \; f\in V\right\}
\; \subset 2^{\cX}, 
\label{def_cB}
\ee
which is the totality of the level sets $ f^{-1} (\lambda) = 
\{ x \,|\, f(x) = \lambda\} \subset \cX$ of functions in $V$.  We first show that, 
for any $B\subset \cX$, 
\be
B \in \cB \; \Leftrightarrow \; 1_B \in V.
\label{1_BinV}
\ee
Since $\Leftarrow$ is obvious, it suffices to show $\Rightarrow$. 
Assume $B \in \cB $, so that  $B = f^{-1} (\lambda)$ for some $f\in V$ and $\lambda\in \bR$. When $B$ is the empty set $\phi$, we have $1_B = 0 \in V$.  So we assume 
$B\neq \phi$, which means that $\lambda\in f(\cX)$.  Let the elements of $f(\cX)$ be 
$\lambda_0, \lambda_1, \ldots \lambda_k$, where $\lambda_0 = \lambda$ and 
$\lambda_i \neq \lambda_j$ if $i\neq j$, and let  $B_i := f^{-1} (\lambda_i)$. Then 
we have  $f = \sum_{i=0}^k \lambda_j 1_{B_i}$ with $B_0 = B$.  
Let $a(t) = a_0 t^k + a_1 t^{k-1} + \cdots + a_k $ be a polynomial 
satisfying $a(\lambda_0) = 1$ and $a(\lambda_i) = 0$ for any $i\neq 0$. 
Explicitly, $a(t)$ is expressed as
\[ 
a(t) = \prod_{i=1}^k \frac{t - \lambda_i}{\lambda_0 - \lambda_i}.
\]
It follows that
\[  a(f) = \sum_{i=0}^k a(\lambda_i) 1_{B_i} = 1_{B_0} = 1_{B}. 
\]
In addition, $a(f) = a_0 f^k + a_1 f^{k-1} + \cdots + a_k $ belongs to 
$V$ since $V$ is a subalgebra of $\bR^{\cX}$ with $1\in V$. 
Hence we have $1_B\in V$. 

Using \eqref{1_BinV}, we see that
\begin{gather}
\cX \in \cB
\label{cXinB},  \\
B\in \cB \; \Rightarrow \; B^c \in \cB,  
\label{Bc} \\
B_1, B_2 \in \cB  \; \Rightarrow \;  B_1 \cap B_2 \in \cB
\label{B1capB2} 
\end{gather}
as 
\begin{align}
1_\cX = 1 \in V \Rightarrow &  \cX \in \cB,  \\
B\in \cB \Rightarrow 1_B\in V \Rightarrow 1_{B^c} =& 1 - 1_B \in V 
\nonumber \\
& \Rightarrow B^c \in \cB, 
\\ 
B_1, B_2 \in \cB \Rightarrow 1_{B_1}, 1_{B_2} \in V & \Rightarrow 
1_{B_1\cap B_2} = 1_{B_1} 1_{B_2} \in V \nonumber \\
& \Rightarrow B_1\cap B_2 \in \cB.
\end{align}

Properties \eqref{cXinB}-\eqref{B1capB2} implies that 
$\cB$ is an additive class of sets ($\sigma$-algebra) on 
the finite entire set $\cX$. Therefore,  $\cB$ is generated by a partition 
$\{A_1 , \cdots , A_n\}$ of $\cX$ in the sense that every element of 
$\cB$  is the union of some (or no) elements of $\{A_1 , \cdots , A_n\}$. 
Recalling the definition \eqref{def_cB} of $\cB$, we conclude 
\eqref{V_partition}. 

\end{proof}

\begin{thebibliography}{1}
\bibitem{Chentsov}
N. N. {\v{C}encov (Chentsov)}, 
{\em Statistical Decision Rules and Optimal Inference}, 
AMS, 1982
(original Russian edition: Nauka, Moscow, 1972).
\bibitem{Amari:CurvedExp82}
S. Amari, 
``Differential geometry of curved exponential families---curvature and
  information loss", 
{\em The Annals of Statistics}, 10, 357--385, 1982.
\bibitem{Amari:springer}
S. Amari, {\it Differential-Geometrical Methods in Statistics}, Springer, Lecture Notes in Statistics 28, 1985.
\bibitem{AN} S. Amari and H. Nagaoka, {\it Methods of information geometry}, AMS \& OUP, 2000.
\bibitem{NA} H. Nagaoka and S. Amari, ``Differential geometry of smooth families of probability distributions", 
Technical Report METR 82-7, Dept. of Math. Eng. and Instr. Phys, Univ. of Tokyo, 
1982. (http://www.keisu.t.u-tokyo.ac.jp/research/techrep/data/1982/METR82-07.pdf)
\bibitem{Ohara2} A. Ohara, 
``Information geometric analysis of an interior point method for semidefinite program- ming",  {\em Geometry in Present Day Science}  (eds. O. E. Barndorff-Nielsen and E. B. V. Jensen), pp.49-74, World Scientific, 1999.
\bibitem{Ohara3} A. Ohara, 
``Geodesics for dual connections and means on symmetric cones", {\em Integr.\ equ.\ oper.\ theory}, 50, 537--548, 2004.
\end{thebibliography}
\end{document}